\documentclass[letterpaper, 10 pt, conference]{IEEEtran}
\IEEEoverridecommandlockouts
\usepackage{graphicx, xcolor, tabularx, booktabs}

\title{\LARGE \bf
Unconventional Exchange: Methods for Statistical\\ Analysis of Virtual Goods
\thanks{This work was supported by the EPSRC Centre for Doctoral Training in Intelligent Games \& Games Intelligence (IGGI) [EP/L015846/1] and the Digital Creativity Labs (digitalcreativity.ac.uk), jointly funded by EPSRC/AHRC/Innovate UK under grant no. EP/M023265/1.}}
\newcommand{\datadate}{09/12/2018}

\author{
\IEEEauthorblockN{Oliver James Scholten\IEEEauthorrefmark{1}, Peter Cowling\IEEEauthorrefmark{1}, Kenneth A. Hawick\IEEEauthorrefmark{2} and James Alfred Walker\IEEEauthorrefmark{1},~\IEEEmembership{Senior Member, IEEE}}\\
\IEEEauthorblockA{\IEEEauthorrefmark{1}Department of Computer Science, University of York, Heslington, York, YO10 5GE, UK \\
Email: \{ojs524, peter.cowling, james.walker\}@york.ac.uk}
\IEEEauthorblockA{\IEEEauthorrefmark{2}Department of Computer Science, University of Hull, Hull, HU6 7RX, UK \\
Email: k.a.hawick@hull.ac.uk}
}

\usepackage[bookmarks=false]{hyperref}
\usepackage{cite}
\hypersetup{linkbordercolor=green}

\begin{document}

\newcommand{\game}{\textit{Old School Runescape}}
\newcommand{\addcite}{\color{red}\textbf{[cite]}\color{black}}
\IEEEoverridecommandlockouts
\IEEEpubid{\begin{minipage}{\textwidth}\ \\[12pt]
978-1-7281-1884-0/19/\$31.00 \copyright 2019 IEEE
\end{minipage}}

\maketitle

\begin{abstract}
Hyperinflation and price volatility in virtual economies has the potential to reduce player satisfaction and decrease developer revenue.
This paper describes intuitive analytical methods for monitoring volatility and inflation in virtual economies, with worked examples on the increasingly popular multiplayer game \game. 
Analytical methods drawn from mainstream financial literature are outlined and applied in order to present a high level overview of virtual economic activity of 3467 price series over 180 trading days. 
Six-monthly volume data for the top 100 most traded items is also used both for monitoring and value estimation, giving a conservative estimate of exchange trading volume of over \pounds60m in real value. Our worked examples show results from a well functioning virtual economy to act as a benchmark for future work.
This work contributes to the growing field of virtual economics and game development, describing how data transformations and statistical tests can be used to improve virtual economic design and analysis, with applications in real-time monitoring systems.
\end{abstract}

\vspace{1mm}

\section{INTRODUCTION}
Virtual hyperinflation reduces individuals ability to purchase items with the virtual currency they have, in effect altering the rate of progression in games such that new players will progress more slowly, and the relative value of existing player's items will decrease over time\cite{Earle2018}.
Hyperinflation causes in virtual economies include incorrect calibration of the mechanisms of addition/deletion\cite{lehdonvirta2014virtual} of virtual currency and/or items, duplication bugs in code, or external factors such as gold farming activity - each increasing the amount of virtual currency in circulation at a rate greater than the economy can comfortably handle.
This paper explores analytical methods which can be used on existing data sets to give virtual economy controllers the ability to identify non-standard behaviour and potentially damaging effects of actions like game updates or micro-transaction additions.
We begin by introducing virtual economies, virtual goods, and the emergent economic phenomena found to take place in virtual worlds.

Virtual economies can be succinctly described as the emergent phenomena of users generating, trading, and consuming virtual goods\cite{Girvan2018}.
This definition is supported by Nazir and Lui's comprehensive review\cite{Nazir2016} and Knowles and Castronova's chapter\cite{knowles2016economics} in the \textit{Handbook on the Economics of the Internet}\cite{Bauer2016}.
These economies have been found to share many similarities with real economies but with several important differences.
These can be summarized as a lack of regulatory oversight, no upper limit on the generation of new virtual assets, and complete dependence for governance on the developer of virtual world.
This freedom gives developers a rich landscape to create impressive user experiences, with games like \game{ }  offering an environment in which players can explore, trade, and increase their avatar's abilities to achieve greater feats.

Virtual goods can be described as a rivalrous digital token in a virtual world, this follows Hamari and Keronen's work\cite{hamari2016people} which investigates why people purchase these virtual goods.
It should be noted that although the creation of the goods in the literal sense, as the data in the database of the developer, requires little effort, it is the creation of the goods under the rules of the virtual world that requires some amount of work (for example defeating some boss or travelling some distance).
With virtual economies and virtual goods briefly introduced, we can discuss the phenomena they exhibit in a general sense, specifically price behaviours caused by developers actions.

As virtual goods themselves can be created and destroyed at will by the controller of the virtual economy, their supply and demand is subject to several important effects.
The introduction of a new virtual good to an existing system, in the form of a game update or new release, is of primary concern.
Upon introduction a new virtual good may devalue existing goods with which it shares similar function, it may grow itself in price given its relative lack of abundance, or it may not move in price at all should in-game vendors offer it at a fixed and replenishable rate.
It is these types of effects which also contribute to virtual economic instability, increasing risks of virtual hyperinflation and other undesirable behaviours like excessive gold farming by players.

Research into virtual economies and their behaviour also has wider effects than game design and player satisfaction.
Work by Castronova\cite{castronova2016players}, Knowles\cite{knowles2015virtual}, and Lehdonvirta\cite{lehdonvirta2010new}, has investigated virtual worlds as the future of low-skilled work, their origins and issues, and their regulation, respectively.
These works all view virtual worlds through serious academic and regulatory lenses to identify issues in their governance and wider impact.
This is especially relevant given the real-value of the currencies and goods existing in virtual worlds as presented in this paper.


Finally, this paper may be outlined as follows: section \ref{sec:ve} discusses the mechanics by which virtual goods are exchanged, and the data gathered as part of this work in Section \ref{sec:data}. 
We then discuss the data analysis methods used in Section \ref{sec:analysis}, providing brief explanations and justifications for each.
With data and methods defined, we present several preliminary findings which are available through pure observation in Section \ref{sec:prelim}. 
We then present the main results including figures describing each of the methods presented in Section \ref{sec:results}.
The discussion and conclusion then bring together the findings and summarize the contribution in Section \ref{sec:dis}, to be followed by a brief critique of our implementations and recommendations for further work in Section \ref{sec:cfw}.

\section{VIRTUAL EXCHANGES}
\label{sec:ve}
With virtual economies, goods, and emergent economic phenomena briefly introduced, we move to outline virtual economic exchanges - the vessel with which any economic instability may be detected.

With more than 100,000 concurrent players\footnote{\url{http://www.misplaceditems.com/rs_tools/graph/}} \game is one of the largest multiplayer role-playing games globally, with over 250 million registered accounts. 
Gameplay revolves around increasing skills, gathering items, and socializing with other players. 
Many of the items each player collects may also be traded on the Grand Exchange (GE) - a limit order exchange with few constraints other than an four-hourly buy volume limit which varies per item. 
This exchange is freely accessible to all players with member (paid subscription) and non-member items combined. 
Bid-Ask spreads (charts showing how much other people are willing to offer/pay for goods) on the GE are not visible to players and cannot be used either for trading or analysis.
Transactions then occur when buyer orders are matched with sellers, or trades are made independently from the exchange.
These direct trades are not accounted for in the data gathered but represent an interesting avenue of further work.

The limited information the GE offers to those using it for trading applies a hard limit to the amount a given player can know about the state of the system as a whole. 
This reduces the likelihood of informed trading of any kind, resulting in a more even playing field for trading parties. 
This detail means models of this particular system may benefit from homogeneous representation of players, a direct contrast to sentiment in the financial literature\cite{voit2013statistical}, and is an key insight for future work.

\section{RUNESCAPE DATA}
\label{sec:data}
Data gathered as part of this work include daily price histories of 3467 items traded on the GE, for the past 180 consecutive days up to \datadate. 
The time-zone independence of the GE means six months of real time are represented as weekends are not excluded. 
Data was gathered through Jagex' publicly accessible GE API, and items with zero price movement over the entire length of the series were excluded (109 items total), leaving 3358 series of interest.

Volume data for the 100 most traded items was also gathered through the Jagex GE webpage\footnote{http://services.runescape.com/m=itemdb\_oldschool/top100}. 
This is used in index construction and for estimating exchange trading volume. 
Volume data for the remaining 3258 items is not available through the GE API so is not included.

\subsection{Transformations}
In order to explore virtual goods price histories effectively a number of transforms may be applied, these allow slight variations of analytical interpretations from the original data set, and allow a more complete analytical picture to be drawn.

\begin{equation}
    I(c) = {\frac{B_{r}}{B_{v}}}c
    \label{eq:realvalue}
\end{equation}

Framing discussion of virtual items with respect to their real world value requires a formal mapping as shown in \ref{eq:realvalue} - a na\"{i}ve estimation of the real-world price of a virtual asset $I$ given it's virtual price $c$ based on the real and virtual prices of in-game bonds ($B_{r}$ and $B_{v}$). 
These are items redeemable for membership by players and can be traded on the GE.
This equation may be generalized for any world by substituting the bond prices for any virtual token officially sold by the developer and it's corresponding in-game price. 
Analytical insights drawn from real-value series following such transformation will not be inherently informative but can be used to present findings on a more intuitive scale.
Note that this may be complicated further by the use of multiple virtual currencies within a world yet this addition only wraps the relationship in layers of conversions so is omitted.

Grey markets for real money trading is a further means to buy and sell in-game currency\cite{Knowles2015}, albeit explicitly against the terms of use for many developers. 
Accounting for prices in such markets remains difficult, hence interpretation of the above equation as-is generalizes to `the price of item $I$ equals the amount of real currency required to purchase it's in-game worth \textit{through official channels}' which may be inflated from it's real sale price, or street value.

Finally, transformations such as taking the logarithm of each value in the series or applying a first differencing filter to the series can be done to enforce stationarity - a key concept in time series analysis.
First differencing simply looks at the differences between values in time as opposed to their value alone, giving better insight into how the underlying system behaves over time.
Taking the natural logarithm is useful both for scaling down large values and approximate estimation of percentage differences between values.

\section{DATA ANALYSIS METHODS}
\label{sec:analysis}
The data analysis methods chosen in this work aim to provide the reader with a battery of methods which may be used to assess the overall health of the virtual economic system in question.
In order to fully explore the data, bivariate plots, price indexes, and stationarity tests following transformations may all be applied, justifications for which are discussed below. 
These may be complimented by fitting distributions to return data, although this does not contribute much to intuitive understanding\cite{Chu2015}. 
Application of this battery of methods means, analytically speaking, that the most complete picture possible can be created, giving designers the best possible chance of improving their virtual economy and finding any anomalous or counter-intuitive behaviours in their systems.

\begin{table*}[]
\caption{Summary of methods, data required, and analytical purpose.}\label{tab:methods}
\centering\normalsize
\begin{tabularx}{16.7cm}{p{3.5cm} p{5cm} p{7cm}}
  \toprule
   Method & Input & Purpose \\ \midrule
   Bivariate Plot & Descriptive statistics for all series & Provide high level overview of price behaviours in the virtual economy \\ \midrule
   Quartile Price Indexes & All series data divided into four groups & Assess 'cost of living' for players in each wealth bracket \\ \midrule
   Top 100 Price Index & Series data of the top 100 most traded items & Assess changes in 'cost of living' for the most popular items \\ \midrule
   Stationarity Tests & Quartile and Top 100 price index returns & Identify any abnormal behaviour in any of the above indexes \\ \bottomrule
\end{tabularx}
\end{table*}

\subsection{Bivariate Plots}
The first step is to create bivariate plots with macroscopic variables of interest. 
Here they include the mean daily percentage change, the virtual price (logarithmically transformed), and the coefficient of variation $c_{v} = \sigma / \mu$ (standard deviation over mean). 
The price is logarithmically transformed such that extreme values may be compared on the same axis, and the coefficient of variation provides a numerically comparable representation of volatility irrespective of the price of a virtual good ($c_{v}$ may be compared between goods of different values, e.g. gold and iron).
In combination these plots provide an intuitive overview of the system, along with mean values of each variable to be compared across data sub-sets. 
Plotting the mean of each variable on each axis also allows approximation of quartile based clusters should they appear.

Other descriptive statistics may be included in subsequent bivariate plots, these may include mean trading volume, number available for trading, and rate of addition into the virtual world (gathering rate). 
Data for such variables was not available through the GE API so is not included in this text but remains a promising area of future work.

\subsection{Price Indexes}
Following bivariate plots, individual indexes should be constructed based on some division of the available goods. 
This decision is somewhat arbitrary but can be used to inform economic balancing and design issues, for example; taking four quantiles and creating indexes for each would allow estimation of inflation in each quarter of virtual goods within the world. 
This quantile based estimation is useful for capturing economic effects in different sub-sections of the virtual economy, giving more granular detail than a single overall estimation, whilst providing more generalisable findings than per-item estimation.
It can safely be assumed that the average player will interact in some way with goods in the lowest quartile - especially new users - hence it is of primary interest. 
Conversely the upper-most quantile likely contains the most advanced late-game items, so is of interest in terms of late-game progression and gauging the impact of new late game content to the world.

Analysis of price indexes not weighted by volume may not give accurate results relative to the `true' behaviour of the system in question. 
This can be described in simple terms as the combined price of `one of everything' in a shop not representing the average customer's purchases which would likely be more milk, bread, and other staple items.
As trading volume data is not explicitly available for all goods a \textit{top 100} index may be created to address the most commonly traded items, similar to stock indexes in the real world\cite{Chen2018}. 
Comparison with other indexes may indicate balancing issues or other design flaws in the system so should be performed part way through the analytical process, these comparisons will become clear in the worked examples.

Indexes may be constructed a number of ways, in this work the price of each item in the quantile in question should be summed yielding a single series, similar to the consumer price index\cite{Nisar2017}. 
In the context of virtual economies a single global index has little value to add in terms of informing design decisions other than monitoring overall inflation, which can then be heavily skewed by newly added items in the upper quartile.

\subsection{Stationarity Tests}
Tests for stationarity can be used to inform whether or not data transformations have successfully removed elements of trends, seasonal effects, and other features not found in purely stochastic series. 
Passing these tests indicates an absence of such elements, such that the transformation used (if any) captures them in their entirety. 
This can be useful in quantifying trends and other features with respect to design decisions in the virtual world in which the virtual economy operates. 
For example we may see regular oscillations at a monthly time window, if these coincided with some monthly promotion by the developer it would be reason to suspect that promotion was having some adverse effect on the virtual economy. 
Here that would mean performing a transformation to remove the monthly trend then testing for stationarity in the remaining series.

Stationarity (specifically difference stationarity) here is tested using the augmented Dickey-Fuller test\cite{Guilkey1989}, with a null hypothesis that unit root exists in the time series under inspection. 
This can be most simply described as a test to identify any non-random behaviours present in a time series (unit root), which may typically be used to identify abnormalities in pricing/trading in market data.
Tests should be performed on each of the indexes following de-trending transformations if applicable. 
These will highlight any abnormal behaviours in each of the series, and when used in conjunction with a regular plot may highlight further avenues of investigation. 
It must be acknowledged that stationarity/unit root tests are just one way of treating time series, for an accessible summary of methods see Chatfield's work\cite{chatfield2016analysis}.

Many more methods exist than could be described in this work, several which may be of interest can be briefly mentioned as potential avenues of further work. 
The methods presented above represent those most applicable to existing virtual economies, and those which may be most intuitively interpreted (see Table \ref{tab:methods}).
Other more sophisticated methods include correlation matrix analysis\cite{Nguyen2018} which can be used to identify similarities in temporal fluctuations between goods, structural equational modelling which aims to formally represent series according to some noise/autocorrelation\cite{Dong2017}, and spectral analysis which aims to decompose series into their component parts/oscillations in the frequency domain\cite{Granger2015}. 
Each of these methods may be used on top of those applied here to give even deeper understanding of the mechanics at play.
We now move to discuss the implementation details of the experiments performed and the tools used for reproducibility.

\subsection{Experiment Details}
The data gathered arrives from the API calls as JSON which is then parsed into data frames according to the Pandas module structure in Python.
With data frames stored, any series showing no movement over the duration are removed, descriptive statistics are then calculated.
Using these statistics the set is then further refined into indexes, specifically quartile indexes as discussed above.
Plots and statistical tests are created and applied here, with the entire process up to this point requiring less than one minute on laptop-level hardware.
The data in its entirety can be comfortably loaded in to memory and analysed, larger sets involving more granular temporal measurements will take more memory but should be achievable on reasonably advanced hardware.
Before full analytical results are calculated the preliminary results requiring little analytical effort can be observed.
The observation of these results brings us to their discussion with respect to our worked example.

\section{PRELIMINARY RESULTS}
\label{sec:prelim}
Several findings stand out in the raw data as being of particular interest with no sophisticated analysis required. These include;
\begin{itemize}
\item The most traded item by volume was the \textit{Fire Rune} with over 25 billion traded in the last 6 months, or \pounds166,250 in real value.
\item Not all items on the exchange have changed value, with 109 showing no movement in the last 180 days.
\item The most expensive item traded is the \textit{Scythe of Vitur}, trading for approximately \pounds1,862.00 each.
\end{itemize}

These insights act to frame further analysis in the context of the virtual world itself and can be used as sanity checks against further findings. 
They might also be used to inform design decisions with respect to player experience, in this case does the \textit{Fire Rune} dominate gameplay at the cost of other items as it does the GE? 
Answers to such questions may be extracted from trading data alone avoiding the need to poll player opinion or other expensive, invasive methods.

\section{RESULTS}
\label{sec:results}
\begin{figure}
  \centering
  \includegraphics[width=\linewidth]{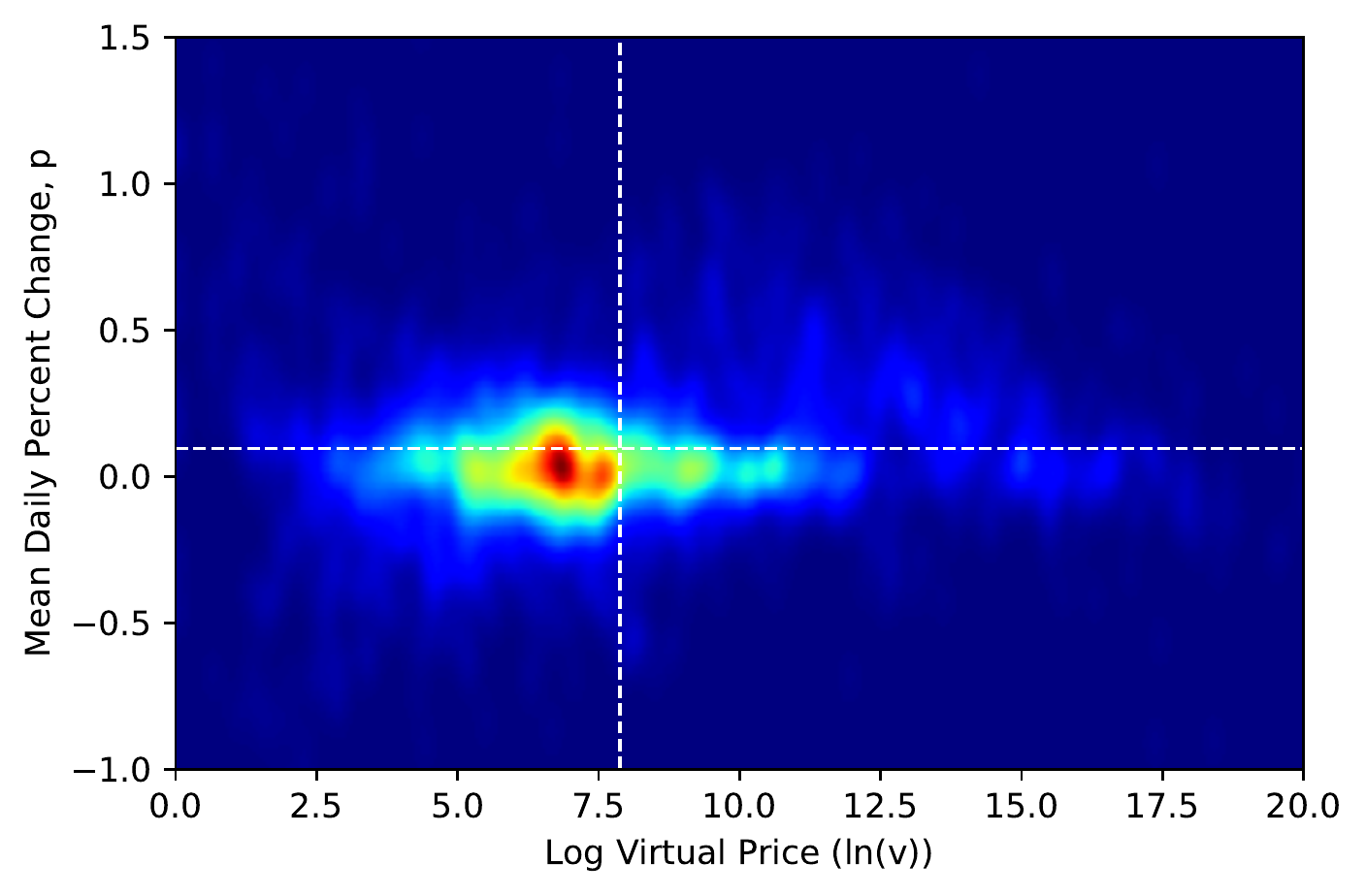}
  \caption{Thermal histogram plot of mean daily change against log virtual price for all goods whose value has changed in the last 180 trading days. Red areas indicate increased density of points, darker blue indicates fewer points in the area. Gaussian blur applied with sigma=8, bins=1000, dotted lines show mean axis values.}
  \label{fig:meanlog}
\end{figure}

\begin{figure}
  \centering
  \includegraphics[width=\linewidth]{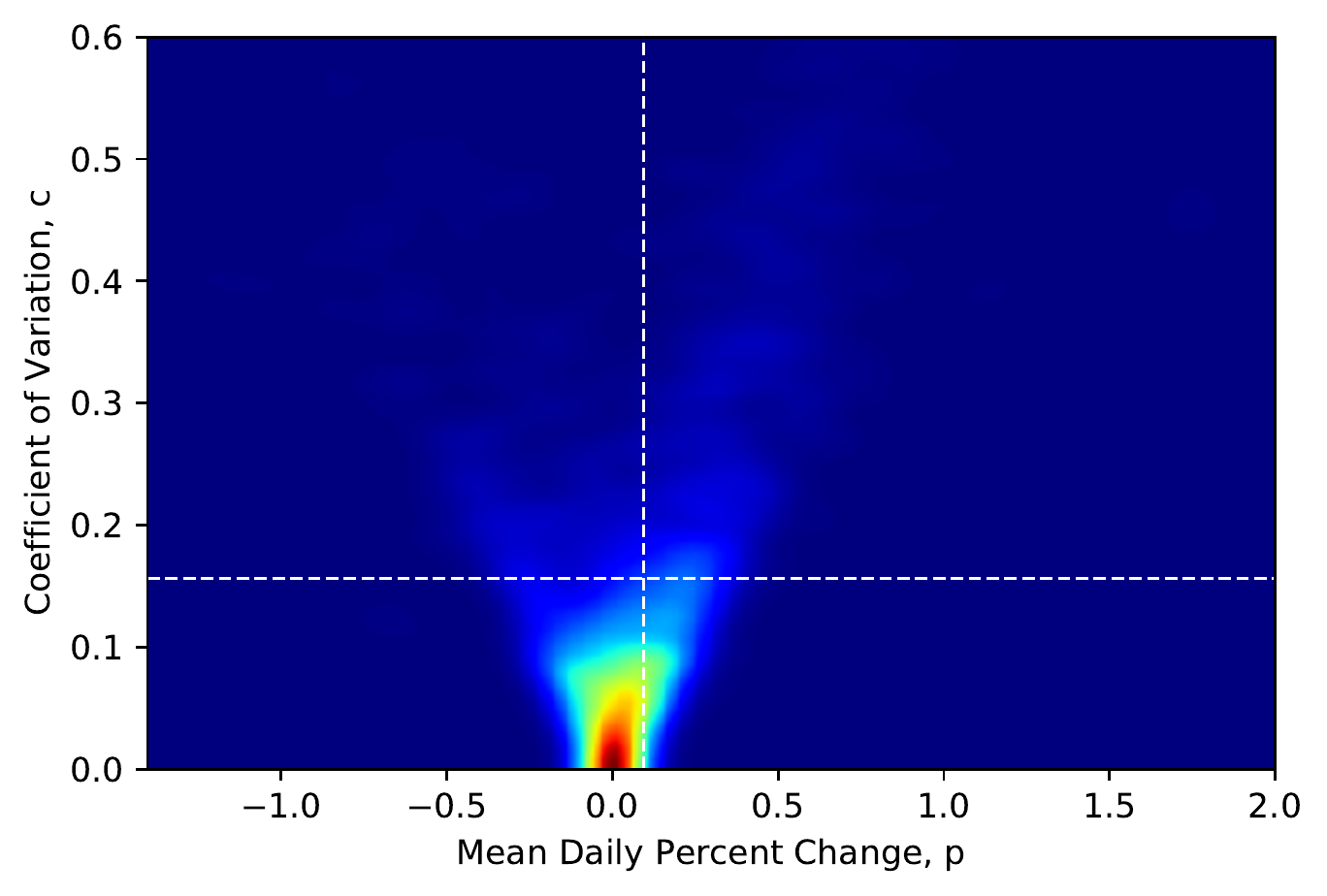}
  \caption{Thermal histogram plot of coefficient of variation against mean daily percentage change for all goods whose value has changed in the last 180 trading days. Red areas indicate increased density of points, darker blue indicates fewer points in the area. Gaussian blur applied with sigma=8, bins=1000, dotted lines show mean axis values.}
  \label{fig:covmean}
\end{figure}

Fig. \ref{fig:meanlog} and Fig. \ref{fig:covmean} show the bivariate plots for each of the 3358 virtual goods of interest. 
Fig. \ref{fig:meanlog} shows a slightly elongated concentration of goods around a mean daily percentage change of zero and a price logarithm of 7.
This tells us that the majority of items in the virtual economy are not increasing or decreasing in value at alarming rates, nor are there any regions of items which go against this norm (there is a single dominant concentration of virtual goods).
From the fairly circular shape of the heatmap we see that approximately as many virtual goods are increasing in value as are decreasing, and that their distribution in terms of their value is also regular.
These findings can be confirmed by the lines indicating the average values for each axis, although their lack of alignment with the center of the cluster indicates some outliers which are both of high value and which are growing.
This can be speculated to be the result of new additions to the economy of high level items.
An unexpected result in this figure may be the existence of multiple peaks, uneven shapes, and any otherwise irregular abnormalities.

\begin{figure}
  \centering
  \includegraphics[width=\linewidth]{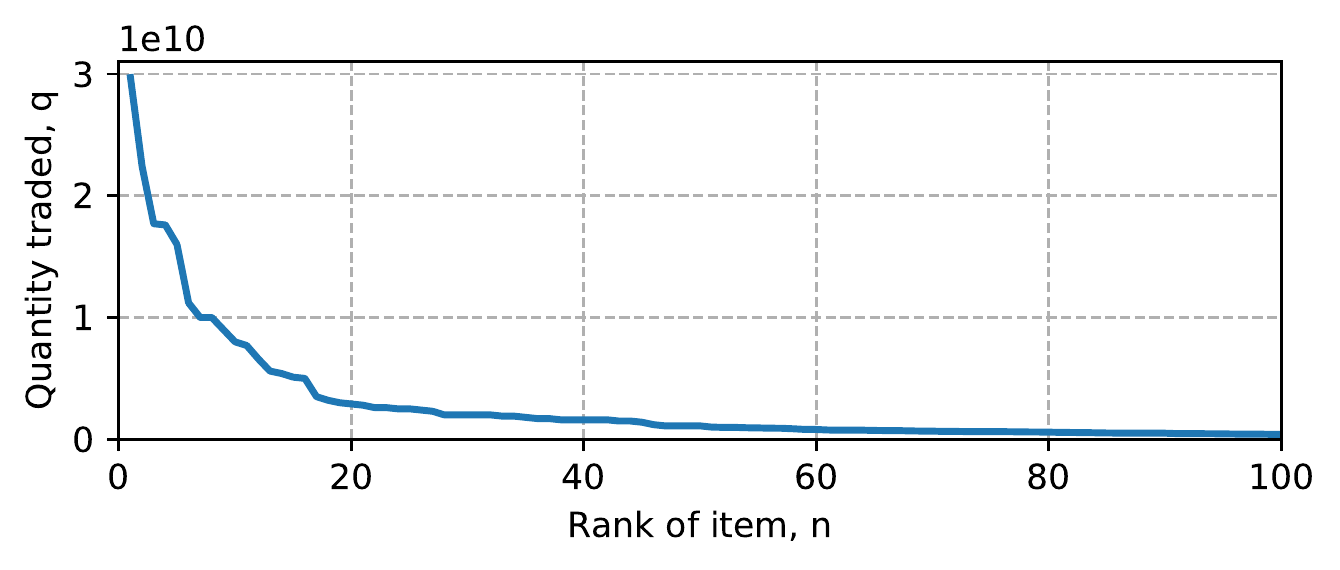}
  \caption{Six month trading volume of each item in the top 100 most traded items.}
  \label{fig:top100}
\end{figure}

Fig. \ref{fig:covmean} shows us a similar style plot instead indicating the volatility of virtual goods with respect to their mean daily percentage change.
This single concentration around zero in the $x$ axis can again be interpreted as a balance in items increasing and decreasing in value, with the addition that the volatility in their movements are also approximately equal.
The lack of skew in the shape of the cluster means that goods are generally behaving identically on each side of zero-growth, signs of imbalance on this chart may manifest as lop-sided clustering or multiple peaks as in the previous figure.

Volume data of the top 100 goods traded can also be plotted as-is to give an intuitive glance at the distribution of items with respect to their volume. 
Fig. \ref{fig:top100} shows this data in its raw form, presenting a Pareto style distribution with exceptionally high volumes in the top 20 items. 
We find that the top 20 items account for 69.7\% of the total volume of items traded in the top 100 data. We may speculate that were the remaining 3258 volumes accounted for, this portion may be higher. 

Using equation \ref{eq:realvalue} we find the real-world value of the volume of the top 20 most traded items - calculated using the mean price multiplied by the total volume - to be approximately \pounds21.5M, with the top 100 summing an impressive \pounds65.6M. 
This finding suggests that the distribution in real-world value of the most traded items doesn't match the distribution of quantity traded. 
This means that the real-world value of the trading history of an item plays a key role in its analytical significance, and that from a design perspective such differences should be examined at the micro-level to ensure such highly traded items are balanced with respect to their distribution in the player base.
The exact reason for this difference in distribution is unknown but represents an interesting area of future work.

\begin{figure}
  \centering
  \includegraphics[width=\linewidth]{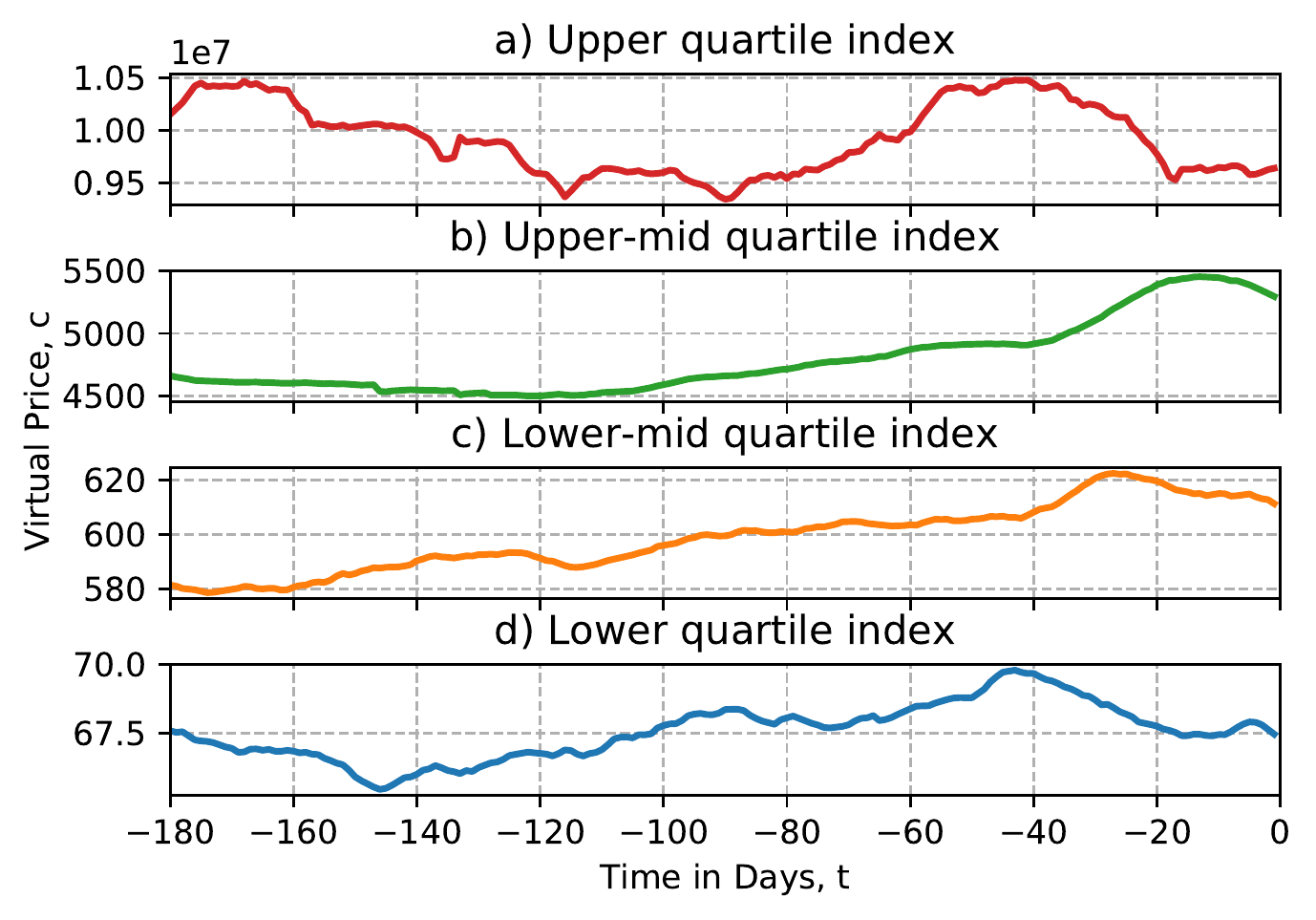}
  \caption{Price indexes for each quartile, upper in red, upper-mid in green, lower-mid in orange, and lower in blue. Note that virtual prices have been scaled down by $10^3$, and that the 1e7 scale applies only to the upper quartile y-axis scale.}
  \label{fig:price_indexes}
\end{figure}

With macroscopic findings presented, we move to construction of the indexes. In this case, quartile indexes are used as they are most intuitively interpreted. Fig. \ref{fig:price_indexes} shows each of the indexes in their raw form, with the upper quartile in red and the lower in blue. 
As with the bivariate plots they appear fairly standard, with no obvious seasonal effects, although the two mid-quartile plots both show a slight upward trend. 
Price indexes allow us to determine inflation rate of sections of the economy, this calculation is simply the difference between the start and end point, divided by the end point. 
This gives us inflation rates presented in Table \ref{tab:index_stats} - which have already been converted to percentages.

It is immediately obvious that inflation rates between indexes are dissimilar in both size and direction, with an average value of 2.85\%, and the upper-mid quartile's growth being the largest of the group. 
This is of direct interest to those responsible for maintaining the stability of this market as it may reflect design decisions regarding the addition of new items or deletion of existing items from these quartiles.
As these figures are for the duration of the gathered data we find a $\approx$12\% six-monthly inflation of the upper-mid quartile. 
This value can be compared to the 50\% over a given time period typically used to define hyperinflation\cite{Cagan1956}.
Hence this value is high in comparison to real world economies but does not constitute hyperinflation, it should however be monitored closely as sustained periods of high inflation in virtual economies can negatively impact the ability of new (or poorer) players to make purchases in that price bracket. 
This in turn may affect player's motivations to participate in micro-transactions and other developer revenue streams although quantifying this remains an area of future work.

The remaining indexes show visual properties typical of stochastic processes, this means that the market can generally be described as functioning normally, with no \textit{statistically perceivable} nefarious behaviours which may manifest as artificial price increases or prolonged fixed values.

\begin{figure}
  \centering
  \includegraphics[width=\linewidth]{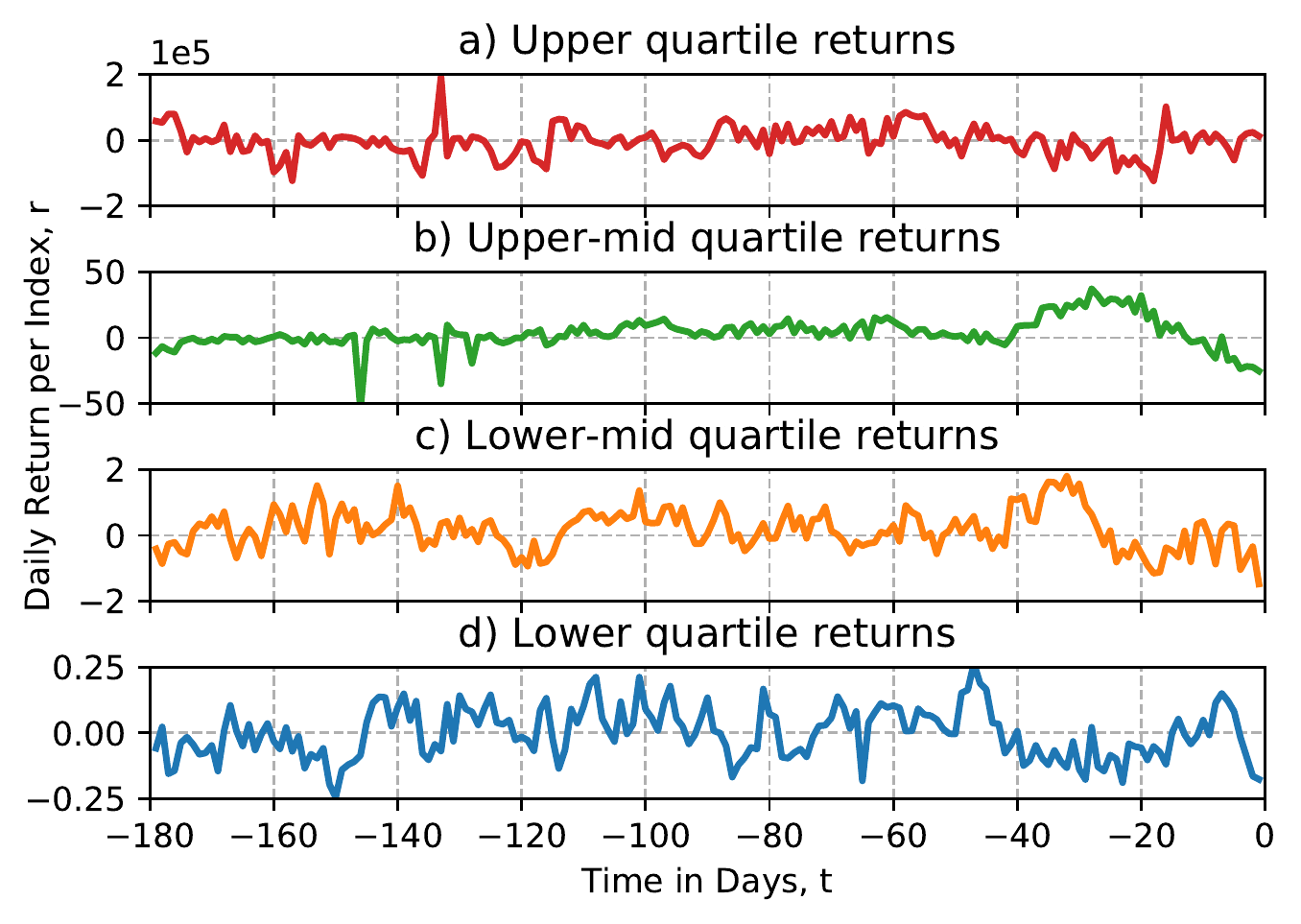}
  \caption{Daily virtual price returns for each quantile index following the same colour scheme and scale adjustment as figure \ref{fig:price_indexes}, the 1e5 scale here again applies only to the upper quartile y-axis.}
  \label{fig:index_returns}
\end{figure}

Fig. \ref{fig:index_returns} shows the first-differenced returns of each of the indexes presented in Fig. \ref{fig:price_indexes}. 
As with the price indexes, subtle differences exist between them, specifically the comparative smoothness of the upper-mid daily returns. This difference can be quantified using the augmented Dickey Fuller test as described in the previous section, the results of which can be found in the second and third columns of Table \ref{tab:index_stats}.

The results of the stationarity tests on the returns of the price indexes are mostly conclusive, with cause to fail to reject the unit root null hypothesis at the 5\% level in all cases, and at the 1\% level in three cases. 
This result if taken alone may constitute cause for concern, specifically regarding behavioural discrepancies in the upper-mid quartile. However, we must recount that construction of these indexes was based on arbitrary intervals, hence this result in the broader context raises no alarms, although further investigation may be fruitful.

\begin{figure}
  \centering
  \includegraphics[width=\linewidth]{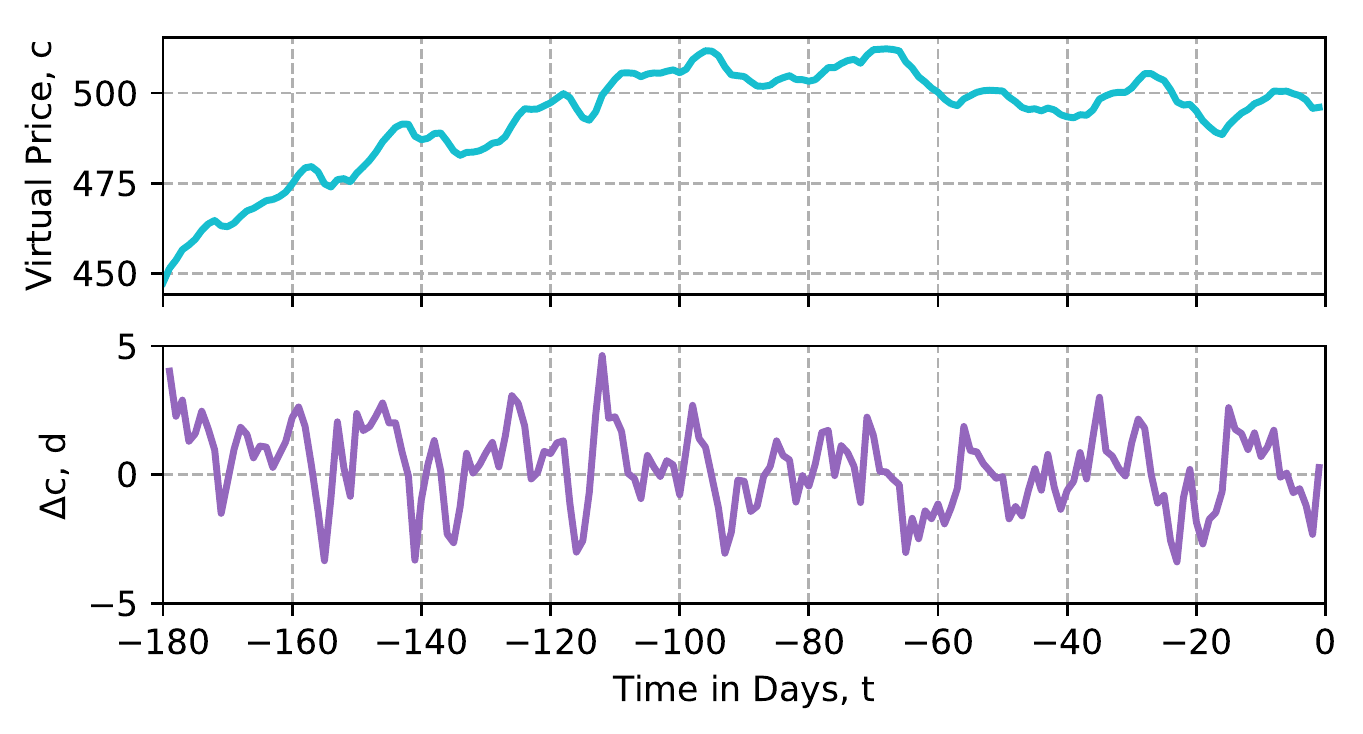}
  \caption{Weighted index constructed using the top 100 most traded items and the first differenced series, weighed using their respective trading volumes. Both y axes have been scaled down by a factor of $10^{11}$.}
  \label{fig:top100_index}
\end{figure}

Finally, Fig. \ref{fig:top100_index} shows both the index of the top 100 most traded items weighted by trading volume, and the first differenced series. Table \ref{tab:index_stats} shows the rate of inflation and stationarity test results respectively.

The weighted top 100 index of the GE appears again to have no seasonal trends, odd behaviours, or otherwise unexpected visual features. This is confirmed by the result of the augmented Dickey-Fuller test statistic with a p-value of 0.007 resulting in failure to reject the null hypothesis at the 1\% level. 
The inflation rate of the top 100 index of 9.81\% is over three times higher than the average of the quantile indexes, this confirms the initial discussion on price index construction that the trading volume is an important factor in analytical results.

Of all of the indexes presented in this paper the top 100 most traded items in figure \ref{fig:top100_index} are assumed to be the closest to the 'true' value of the global weighted index. 
This difference in construction between weighted and un-weighted with respect to trade volume had significant impact on the applicability of the index, but little impact on the analytical results.

\begin{table}[]
\caption{First-differenced index test results and inflation rates.}\label{tab:index_stats}
\centering\normalsize
\begin{tabular}{ @{} l  r  r  r @{} }
  \toprule
   Index Quartile    & ADF $t$-value     & $p$-value & Inflation (\%)    \\ \midrule
   Lower         &  -6.332        & $< 0.001 $   & -0.18      \\ 
   Lower-mid  & -4.781       &  0.046    & 4.91   \\ 
   Upper-mid         & -2.891        & $< 0.001$     &  11.97  \\ 
   Upper        &  -6.082        & $< 0.001$  &  -5.30  \\ 
   Top 100        &  -3.563       & 0.007  &  9.81  \\ \bottomrule
\end{tabular}
\end{table}

\section{DISCUSSION \& CONCLUSION}
\label{sec:dis}
The bivariate plots presented in this analysis show a healthy economy based on intuitive interpretation, given the age and relative stability of the economy in \game this is to be expected. 
Both the coefficient of variation and the mean daily percentage change have proven useful as tools to determine the high-level financial state of a virtual economy, each providing specific insight into questions of interest. 
Namely `how volatile are the goods in this system with respect to their long term growth?', and `how stable are the values of goods in this system with respect to their price?'. 
Both questions have significant impacts in the design and balancing of the virtual economies they concern, with implementation of new faucets and drains\cite{lehdonvirta2014virtual} at the center of solving such issues.

The price indexes constructed here show just one way with which a developer can gain insight into the function of their virtual economy, but have proven useful for estimating inflation and detecting healthy behaviours or non-standard effects. 
Other more sophisticated analytical methods have also briefly been discussed, and areas of further work identified as encountered.

The data set chosen to showcase these methods happens to be that of a relatively stable and well-functioning virtual economy, especially given the number of active participants and trading volume. 
Whilst this paper did not discuss the regulatory impacts of findings using such techniques, they play a key part in the wider literature of analysing virtual economies - specifically trading volume calculation as presented here as this represents over \pounds60m of unregulated, untaxed trades\footnote{although taxation of virtual goods trades themselves is still an open debate.}.

This paper presents data gathered from the \game Grand Exchange, and applies analytical methods similar to those used on financial goods in the non-virtual world. 
Several methods have been briefly introduced and interpreted, providing a strong foundation on which to continue exploration in the virtual world.
The use of trading volume data has been included for the top 100 virtual goods, inviting a more complete analysis. However, as complete a picture as possible has been created with the data available.
Findings are directly applicable to game developers, providing insights into the effects of design choices on the stability of in-game economies, and more generally to virtual economies across digital platforms.

\section{CRITIQUE \& FURTHER WORK}
\label{sec:cfw}
The data analysis presented here fails to capture the volume of \textit{all} of the goods being traded, as such the indexes created can only act as a rough approximation for the more commonly used weighted indexes found in the real world. 
This means that although the `true' results of this analysis may elude us, the methods used would be identical to those presented here and can be performed by those with access to the required data.
Structural equational models may also be fit to the top 100 distribution, quantifying an exponential relationship for example may provide more useful metrics with which to compare subsequent economies.

Further work to this paper may take a number of routes, through the lenses of a number of different disciplines. 
Purist economists may continue the strict application of models outlined in previous sections in order to add to our understanding of the differences between the real and the virtual economies. 
Human-computer interaction researchers may continue investigation on the role of virtual economic wealth on player satisfaction and micro-transaction participation. 
Computationally oriented researchers may choose to extend the designs of such systems to automatically account for the types of issues discussed here. 
Each avenue presents an exciting and potentially lucrative area of work, with applications across the digital entertainment industry.


\bibliographystyle{IEEEtran}
\bibliography{OJS}

\end{document}